\documentclass[aps,twocolumn,longbibliography] {revtex4-1}
\usepackage{amsmath,graphicx,amsfonts}

\usepackage{times}
\usepackage[colorlinks=true,linkcolor=blue,urlcolor=blue,citecolor=blue, breaklinks=true,hypertexnames=false]{hyperref}

\usepackage{amssymb}
\usepackage{bbm}

\usepackage[utf8]{inputenc}
\DeclareUnicodeCharacter{03BC}{\mbox{$\mu$}}

\DeclareUnicodeCharacter{03B4}{$\delta$}
\DeclareUnicodeCharacter{2009}{-}

\DeclareMathOperator{\sech}{sech}

\usepackage[english]{babel}
\addto\captionsenglish{}
\addto\captionsenglish{}

\makeatletter
\def\bbl@set@language#1{%
	\edef\languagename{%
		\ifnum\escapechar=\expandafter`\string#1\@empty
		\else\string#1\@empty\fi}%
	\@ifundefined{babel@language@alias@\languagename}{}{%
		\edef\languagename{\@nameuse{babel@language@alias@\languagename}}%
	}%
	\select@language{\languagename}%
	\expandafter\ifx\csname date\languagename\endcsname\relax\else
	\if@filesw
	\protected@write\@auxout{}{\string\select@language{\languagename}}%
	\bbl@for\bbl@tempa\BabelContentsFiles{%
		\addtocontents{\bbl@tempa}{\xstring\select@language{\languagename}}}%
	\bbl@usehooks{write}{}%
	\fi
	\fi}
\newcommand{\DeclareLanguageAlias}[2]{%
	\global\@namedef{babel@language@alias@#1}{#2}%
}
\makeatother

\DeclareLanguageAlias{en}{english}

\makeatletter
\def\@bibdataout@aps{%
	\immediate\write\@bibdataout{%
		@CONTROL{%
			apsrev41Control%
			\longbibliography@sw{%
				,author="08",editor="1",pages="1",title="0",year="1"%
			}{%
				,author="08",editor="1",pages="1",title="",year="1"%
			}%
		}%
	}%
	\if@filesw \immediate \write \@auxout {\string \citation {apsrev41Control}}\fi 
}
\makeatother

\def\C {\scriptscriptstyle{{C}}}
\def\Q {\scriptscriptstyle {Q}}

\date\today

\def\TITLE{The density of a one-dimensional Bose gas far from an impurity}

\begin{document}
	\title{\TITLE}
	\author{Aleksandra Petkovi\'{c}}
	\author{Zoran Ristivojevic}
	\affiliation{Laboratoire de Physique Th\'{e}orique, Universit\'{e} de Toulouse, CNRS, UPS, 31062 Toulouse, France}
	
	\begin{abstract}
		We consider an impurity in a one-dimensional weakly-interacting Bose gas and  analytically calculate the density profile of the Bose gas. Within the mean-field approximation, by increasing the distance from the impurity, the Bose gas density saturates exponentially fast to its mean thermodynamic-limit value at distances beyond the healing length. The effect of quantum fluctuations drastically changes this behavior, leading to a power law decay of the density deviation from the mean density. At distances longer than the healing length and shorter than a new length scale proportional to the impurity coupling strength, the power-law exponent is $2$, while at longest distances the corresponding exponent becomes $3$. The latter crossover does not exist in two special cases. The first one is realized for infinitely strongly coupled impurity; then the density deviation always decays with the exponent $2$. The second special case occurs when the new length scale is smaller than the healing length, i.e., at weak impurity coupling; then the density deviation always decays with the exponent $3$. The obtained results are exact in the impurity coupling strength and account for the leading order in the interaction between the particles of the Bose gas. 
	\end{abstract}
	\maketitle
	
	\section{Introduction}
	
	Impurities play the pronounced role in one-dimensional quantum liquids and determine their properties. For example, a single impurity severely affects the transport in fermionic liquids \cite{kane_transmission_1992}. In these systems the impurity also causes Friedel oscillations of the local particle density whose envelop decays following the power law $1/|x|^K$ at long distances  \cite{fabrizio_interacting_1995,egger_friedel_1995}. Here $|x|$ is the separation from the impurity and $K$ is the Luttinger liquid parameter. 
	
\begin{figure}[th]
		\includegraphics[width=\columnwidth]{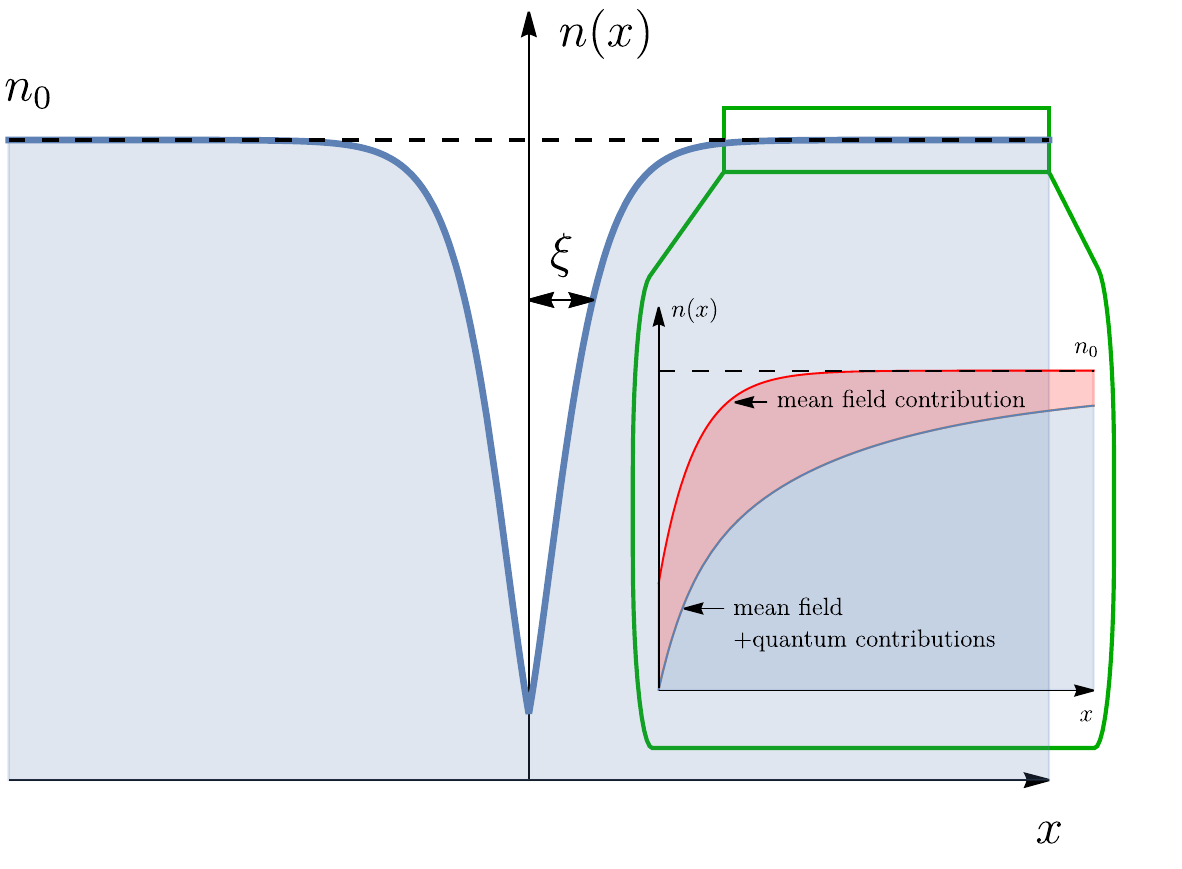}
		\caption{Plot of the local density $n(x)$ of a weakly-interacting Bose gas in the vicinity of the impurity. The inset shows magnified region of the density at  distances from the impurity beyond the healing length $\xi$. It shows the difference between the density in the mean field approximation that very quickly approaches the mean density $n_0$ and the actual density that in addition contains the contribution arising from quantum fluctuations.} \label{fig}
	\end{figure}

	Related question for the behavior of the density in one-dimensional liquids with bosonic particles is less well understood. The goal of this paper is to study this basic problem. Consider the system that consists of a static impurity placed in a weakly interacting Bose gas. In the absence of the impurity, the mean value of the density operator, i.e., the local density, has a constant value. This picture changes once the impurity is present and the density becomes a function of the distance from the impurity. Due to the repulsion from the impurity, the particles of the gas will be pushed away from its position. Since the particles of the medium are not independent, but interact among themselves, the local particle suppression from the region very close to the impurity will impose the particle dilution in the whole  system. This can be visualized as a hole in the Bose gas density created due to the presence of the impurity, see Fig.~\ref{fig}. Our goal is to describe the shape of this hole.
	
	Interacting Bose gas shows superfluid properties. The latter is characterized by a microscopic lengthscale $\xi$ known as the healing length. It generally denotes the distance where the superfluid density recovers from a local perturbation. For our system, the impurity serves as a localized density perturbation and one thus expects the density recovery at separations from the impurity of several $\xi$. This is correct scenario within the mean field picture, which will be discussed below. The corresponding nonlinear equation that determines the density has an exact solution. It shows that the density depletion beyond $\xi$ disappears exponentially fast. In this paper we show that the latter scenario drastically changes beyond the mean-field treatment, i.e., once one accounts for the effect of quantum fluctuations, see Fig.~\ref{fig}. In this case the density depletion decays algebraically with the distance. Formally, $n_0-n(x)\propto 1/|x|^\delta$, where $n_0$ and $n(x)$  respectively denote the mean and the local densities. The exponent  $\delta$ takes the universal values $2$ or $3$ depending on the details such as the impurity coupling strength, which we will be discussed later. Power-law decay of the density depletion means that the presence of the impurity can be probed in the local density far from the impurity. Let us notice that unlike at distances beyond $\xi$ where the quantum contribution prevails, it is parametrically smaller than the mean field contribution in the vicinity of the impurity. In the remaining part of this paper we  derive these results and discuss their further physical implications.

	\section{Theoretical model}
	
	We study a one-dimensional system of bosons with short-range repulsion that is locally coupled to an impurity. The corresponding Hamiltonian is given by
	\begin{align}\label{eq:H}
		\hat H={}&\int_{-\infty}^{\infty} dx\left(-\hat\Psi^\dagger \frac{\hbar^2\partial_x^2}{2m}\hat\Psi+\frac{g}{2}\hat\Psi^\dagger \hat\Psi^\dagger \hat\Psi \hat\Psi\right)\notag\\
		&+G \hat\Psi^\dagger(0,t)\hat \Psi(0,t).
	\end{align}
	Here $m$ denotes the mass of bosons, $g>0$ is the repulsion strength, and $G$ denotes the coupling of the system to an impurity at the origin. The bosonic field operators satisfy the standard commutation relations, $[\hat\Psi(x,t),\hat\Psi^\dagger(y,t)]=\delta(x-y)$ and $[\hat\Psi(x,t),\hat\Psi(y,t)]=0$. We study the system in the thermodynamic limit where the particle density $n_0$ is finite. We are not interested in the impurity dynamics and thus the case when it is infinitely heavy is considered. The impurity serves as a source of a local static potential for the system. For repulsive coupling, $G>0$, the impurity causes the depletion of the Bose gas density at its position in the ground state. Such local perturbation of the density at a single point also affects the local density everywhere in space. The goal of this paper is to calculate how such local depletion will decay as a function of the separation from the impurity. We will calculate the mean boson density
	\begin{align}\label{eq:density}
		n(x)=\langle \hat\Psi^\dagger(x,t) \hat\Psi(x,t)\rangle.
	\end{align}
	Here the average is with respect to the ground state of the system. Since the impurity is static, we do not expect time dependence in the density profile and we thus omitted the time coordinate in the left-hand side of Eq.~(\ref{eq:density}).
	
	We treat the Hamiltonian (\ref{eq:H}) using the Heisenberg equation of motion of the field operator, $i\hbar \partial_t\hat\Psi(x,t)=[\hat\Psi(x,t),\hat H]$. For the purpose of lighter notation, it is convenient to define the dimensionless coordinates $X$ for space   and $T$ for time by 
	\begin{align}\label{eq:X}
		X=\frac{x}{\xi_\mu},\quad T=\frac{t \mu}{\hbar}.
	\end{align}
	Here $\mu$ is the chemical potential of the interacting Bose gas and $\xi_\mu=\hbar/\sqrt{m\mu}$. After introducing the dimensionless field operator $\hat\psi(X,T)$ and the dimensionless coupling $\widetilde G$ by the relations
	\begin{gather}\label{eq:Psipsi}
		\hat\Psi(x,t)=\sqrt{\frac{\mu}{g}}\hat\psi(X,T)e^{-iT},\\
		\widetilde G=\frac{G}{\hbar}\sqrt\frac{m}{\mu},
	\end{gather}
	the equation of motion for $\hat\psi(X,T)$ takes the form
	\begin{align}\label{eq:eom}
		i\partial_T \hat\psi=\left[-\frac{\partial_X^2}{2}+\hat\psi^\dagger\hat\psi-1+\widetilde{G}\delta(X)\right]\hat\psi,
	\end{align}
	while the nontrivial commutation relation is given by
	\begin{align}\label{eq:cr}
		[\hat\psi(X,T),\hat\psi^\dagger(Y,T)]=\alpha^2 \delta(X-Y).
	\end{align}
	Here $\alpha=(\gamma gn_0/\mu)^{1/4}$ is expressed in terms of $\gamma=mg/\hbar^2 n_0$, which denotes the dimensionless parameter of the model (\ref{eq:H}). Up to this point our consideration is general. Equations (\ref{eq:eom}) and (\ref{eq:cr}) follow exactly from the Hamiltonian (\ref{eq:H}) and the initial commutation relations. 
	
	Let us consider the case of weak interaction corresponding to $\gamma\ll 1$. Then the chemical potential is approximately given by $\mu\approx gn_0$ and thus $\alpha\approx \gamma^{1/4}\ll 1$. We then seek for the solution of Eq.~(\ref{eq:eom}) as a series of the form
	\begin{align}\label{eq:psiseries}
		\hat\psi(X,T)=\psi_0(X)+\alpha \hat\psi_1(X,T)+\alpha^2\hat\psi_2(X,T)+\cdots.
	\end{align}
	The function $\psi_0(X)$ can be interpreted as the time independent single particle wave function in the absence of quantum fluctuations, whereas $\hat\psi_{1}(X,T)$ and $\hat\psi_{2}(X,T)$ represent, respectively, its first and second quantum correction.

	Substituting Eq.~(\ref{eq:psiseries}) in Eq.~(\ref{eq:eom}) we obtain a hierarchy of equations controlled by the different powers of $\alpha$. At the leading $\alpha^0$ order we obtain
	\begin{align}\label{eq:GP}
		\widehat{\mathcal{L}}_1\psi_0(X)+\widetilde G \delta(X)\psi_0(X)=0,
	\end{align}
	where we have introduced the family of operators
	\begin{align}\label{eq:L}
		\widehat{\mathcal{L}}_j(X)=-\frac{\partial_X^2}{2}+j|\psi_0(X)|^2-1.
	\end{align}
	Note that $\widehat{\mathcal{L}}_j(X)$ can be determined only after the nonlinear equation (\ref{eq:GP}) has been solved. Expression (\ref{eq:GP}) is known as the Gross-Pitaevskii equation.  At order $\alpha$ we obtain the equation for the first quantum correction of the field operator,
	\begin{align}\label{eq:psi1}
		\left[i\partial_T -\widehat{\mathcal{L}}_2(X) -\widetilde{G}\delta(X)\right]\hat\psi_1(X,T)=\psi_0(X)^2 \hat\psi_1^\dagger(X,T).
	\end{align}
	Equation~(\ref{eq:psi1}) is linear, but it depends on the solution of the nonlinear problem (\ref{eq:GP}). Finally, at order $\alpha^2$ we obtain the equation for the second quantum correction,
	\begin{align}\label{eq:psi2}
		&\left[i\partial_T -\widehat{\mathcal{L}}_2(X)-\widetilde{G}\delta(X)\right]\hat\psi_2(X,T)=\psi_0(X)^2 \hat\psi_2^\dagger(X,T)\notag\\
		&+ 2\psi_0(X) \hat\psi_1^\dagger(X,T)  \hat\psi_1(X,T) + \psi_0^*(X) \hat\psi_1(X,T)^2 .
	\end{align}

	\section{Solution of the equation of motion}
	
	\subsection{Mean-field solution}
	
	Our goal is to describe the ground state of the system. The corresponding solution of Eq.~(\ref{eq:GP}) does not have nodes at any finite value of $\widetilde G$. We thus need strictly positive (or negative) $\psi_0(X)$ that satisfies $\widehat{\mathcal{L}}_1(X)\psi_0(X)=0$ in the two regions $X<0$ and $X>0$, while at the impurity position the boundary conditions
	\begin{gather}
		\psi_0(0^+)=\psi_0(0^-),\\
		\psi_0'(0^+)-\psi_0'(0^-)=2\widetilde{G}\psi_0(0)
	\end{gather}
	must be satisfied. For the solution we find \cite{Note1,kovrizhin_exact_2001}\footnotetext{Note that Eq.~(\ref{eq:psi0sol}) is also the nodeless solution of Eq.~(\ref{eq:GP}) in the case $\mbox{$\widetilde{G}$}<0$. In this case the solution is nonsingular (normalizable) at any attraction apart from the extreme case $\mbox{$\widetilde{G}$}\to-\infty$.}
	\begin{subequations}\label{eq:psi0sol}
		\begin{align}
			\psi_0(X)=\tanh(|X|+X_0),
		\end{align}	
		where	
		\begin{align}\label{eq:psi0solb}
			\tanh X_0=\frac{2}{\widetilde G+\sqrt{4+\widetilde{G}^2}}.
		\end{align}
	\end{subequations}
	In the case of strongly coupled impurity, $\widetilde G\gg 1$, we obtain $X_0\simeq 1/\widetilde G$. Therefore,  $\psi_0(X)=|\tanh X|$ at $\widetilde{G}\to+\infty$. This is the limiting case where the impurity is so strongly coupled that it completely suppresses the Bose gas density at its position. In the opposite case $\widetilde{G}\ll 1$, the Bose gas density is only slightly perturbed at the impurity position. Indeed, we have $X_0\simeq \ln(2/\sqrt{\widetilde G})$ and thus $\psi_0(X)=1-\widetilde{G}e^{-2|X|}/2+O(\widetilde{G}^2)$.
	
	\subsection{First quantum correction}

	Having obtained $\psi_0(X)$, see Eq.~(\ref{eq:psi0sol}), we are in a position to solve Eq.~(\ref{eq:psi1}). This is an operator equation.  We seek for its solution in the form
	\begin{align}\label{eq:psi1ansatz}
		\hat\psi_1(X,T)=\sum_k N_k\left[u_k(X)\hat b_k e^{-i \epsilon_k T}-v_k^*(X)\hat b_k^\dagger e^{i \epsilon_k T}\right],
	\end{align}
	where the summation is over real $k$ values. In Eq.~(\ref{eq:psi1ansatz}), $N_k$ is a real normalization factor that is dictated by the commutation relation (\ref{eq:cr}), $\hat b_k$ and  $\hat b_k^\dagger$ are the quasiparticle bosonic operators that satisfy the commutation relations $[\hat b_k,\hat b_q^\dagger]=\delta_{k,q}$ and $[\hat b_k,\hat b_q]=0$, while $\epsilon_k$ is the  quasiparticle energy for the momentum $k$. Substituting Eq.~(\ref{eq:psi1ansatz}) in Eq.~(\ref{eq:psi1}), we obtain a linear system of two coupled equations for the complex functions $u_k(X)$ and $v_k(X)$. They are known as the  Bogoliubov-de Gennes equations and take the form
	\begin{subequations}
		\label{eq:bdg}
		\begin{gather}\label{eq:bdg1}
			\left[\widehat{\mathcal{L}}_2(X) +\widetilde{G}\delta(X)-\epsilon_k\right] u_k(X)=\psi_0(X)^2 v_k(X),\\
			\label{eq:bdg2}
			\left[\widehat{\mathcal{L}}_2(X) +\widetilde{G}\delta(X)+\epsilon_k\right] v_k(X)=\psi_0(X)^2 u_k(X).
		\end{gather}
	\end{subequations}
	Here, $	\widehat{\mathcal{L}}_j(X)$ operators are defined by Eq.~(\ref{eq:L}) supplemented by $\psi_0(X)$ of Eq.~(\ref{eq:psi0sol}). The system (\ref{eq:bdg}) can be expressed in another form. It is given by
	\begin{subequations}
		\label{eq:upmv}
		\begin{gather}\label{eq:upmv1}
			\widehat{\mathcal{L}}_1(X) \left[u_k(X)+v_k(X)\right]=\epsilon_k \left[u_k(X)-v_k(X)\right],\\
			\label{eq:upmv2}
			\widehat{\mathcal{L}}_3(X) \left[u_k(X)-v_k(X)\right]=\epsilon_k \left[u_k(X)+v_k(X)\right],
		\end{gather}
	\end{subequations}
	and should be considered in the two regions, $-\infty<X<0$ and $0<X<+\infty$, and  supplemented by the boundary conditions
	\begin{subequations}
		\label{eq:uvconditions}
		\begin{gather}
			u_k(0^+)=u_k(0^-),\\
			v_k(0^+)=v_k(0^-),\\
			u_k'(0^+)- u_k'(0^-)=2\widetilde G u_k(0),\\
			v_k'(0^+)- v_k'(0^-)=2\widetilde G v_k(0).
		\end{gather}
	\end{subequations}
	Therefore, the functions $u_k(X)$ and $v_k(X)$ entering $\hat\psi_1(X,T)$ as in Eq.~(\ref{eq:psi1ansatz}) are continuous with a finite jump in its first derivative at the impurity position. Note that in the case of infinite $\widetilde{G}$, the conditions (\ref{eq:uvconditions}) are replaced by $u_k(0)=v_k(0)=0$. This arises from  $\hat\psi(0,T)=0$, which denotes the complete suppression of the boson density at the impurity position.
	
	The solution of the system (\ref{eq:upmv}) supplemented by the boundary conditions (\ref{eq:uvconditions}) was achieved in Refs.~\cite{kovrizhin_exact_2001,petkovic_local_2022}. Here we briefly comment the main steps and give the final results. Combining the two equations of the system (\ref{eq:upmv}), we obtain 
	\begin{align}\label{eq:u+v}
		\widehat{\mathcal{L}}_3(X)\widehat{\mathcal{L}}_1(X) \left[u_k(X)+v_k(X)\right]=\epsilon_k^2 \left[u_k(X)+v_k(X)\right].
	\end{align}
	Once determined the sum of $u_k(X)$ and $v_k(X)$ by solving Eq.~(\ref{eq:u+v}), its difference follows from Eq~(\ref{eq:upmv1}).
	The fourth-order homogeneous differential equation  (\ref{eq:u+v}) has four independent particular solutions, which impose the energy dispersion relation of the Bogoliubov form,
	\begin{align}\label{eq:BS}
		\epsilon_k=\sqrt{k^2+k^4/4}.
	\end{align}
	Forming linear combinations of the four particular solutions one can find the general solution that satisfies the boundary conditions (\ref{eq:uvconditions}). It is given by
	\begin{widetext}
		\begin{align}\label{eq1}
			u_k(X)=\left(1+\frac{k^2}{2\epsilon_k}\right)[f(X,ik)+r(k) f(X,-ik)] +\left(1-\frac{q^2}{2\epsilon_k}\right)r_e(k) f(X,q) + \frac{g(X,i k)+r(k) g(X,-ik)+r_e(k) g(X,q)}{\epsilon_k},\\
			\label{eq2}
			v_k(X)=\left(1-\frac{k^2}{2\epsilon_k}\right)[f(X,ik)+r(k) f(X,-ik)] +\left(1+\frac{q^2}{2\epsilon_k}\right)r_e(k) f(X,q) - \frac{g(X,i k)+r(k) g(X,-ik)+r_e(k) g(X,q)}{\epsilon_k},
		\end{align}
		for $X<0$ and
		\begin{align}
			u_k(X)=\left(1+\frac{k^2}{2\epsilon_k}\right)t(k) f(X,ik) +\left(1-\frac{q^2}{2\epsilon_k}\right)r_e(k) f(X,-q) + \frac{t(k) g(X,ik)+r_e(k) g(X,-q)}{\epsilon_k},\\\label{eq4}
			v_k(X)=\left(1-\frac{k^2}{2\epsilon_k}\right)t(k) f(X,ik) +\left(1+\frac{q^2}{2\epsilon_k}\right)r_e(k) f(X,-q) - \frac{t(k) g(X,ik)+r_e(k) g(X,-q)}{\epsilon_k},
		\end{align}
		for $X>0$. Equations (\ref{eq1})--(\ref{eq4}) assume $k>0$. There we have introduced two auxiliary functions
		\begin{gather}
			f(x,y)=\left[y/2-\mathrm{sgn}(x) \tanh(|x|+X_0)\right]e^{yx},\\
			g(x,y)=\frac{ye^{y x}}{2\cosh^2 (|x|+X_0) },
		\end{gather}
		the abbreviations $q=\sqrt{4+k^2}$, $\eta=\tanh X_0$, and \cite{petkovic_local_2022}
		\begin{gather}\label{eq:r}
			r(k)=\frac{i(1-\eta^2)k[k^2(2\eta+q)+4\eta(\eta^2+1)+q^3+4\eta^2 q+2\eta q^2]} {(k-iq)[\eta k+i(\eta^2+1)][k^2(2\eta+q)+i k(2\eta+q)^2-2\eta(2\eta^2+q^2+2\eta q-2)]},\\
			t(k)=r(k)-\frac{i-\eta k+i \eta^2}{i+\eta k+i \eta^2},\\
			r_e(k)=\frac{4\eta(\eta^2-1) k}{(k-iq)[k^2(2\eta+q)+i k(2\eta+q)^2-2\eta(2\eta^2+q^2+2\eta q-2)]}.
		\end{gather}	
	\end{widetext}
	We notice the properties $|r(k)|^2+|t(k)|^2=1$, $r^*(k)=r(-k)$, $t^*(k)=t(-k)$, and $r_e^*(k)=r_e(-k)$. Since our problem has a static impurity at the origin, the system is symmetric under the spatial inversion. Therefore, in addition to the solution given by Eqs.~(\ref{eq1})--(\ref{eq4}) there is another one where the coordinates and the momenta are reversed. It is formally given by
	\begin{align}\label{eq:inversion}
		u_{-k}(X)=u_k(-X),\quad v_{-k}(X)=v_k(-X),
	\end{align}
	where $k>0$. The summation over negative $k$ in $u_k(X)$ and $v_k^*(X)$ of Eq.~(\ref{eq:psi1ansatz}) is thus defined via the relations (\ref{eq:inversion}). 
	
	Equations (\ref{eq1})--(\ref{eq4}) provide the solution for $\hat\psi_1(X,T)$, see Eq.~(\ref{eq:psi1ansatz}), for arbitrary $X$. They greatly simplify for $|X|\gg 1$. The expression for $u_k(X)$ then becomes
	\begin{align}\label{eq21}
		&{}u_k(X)=\left(1+\frac{k^2}{2\epsilon_k}\right)\biggl\{  \biggl[\frac{1-\mathrm{sgn}\;\!X}{2}r(k)-\frac{1+\mathrm{sgn}\;\!X}{2} t(k)\biggr]\notag\\
		& \times\left(1-\frac{i k}{2}\right) e^{i k |X|}+\frac{1-\mathrm{sgn}\;\!X}{2} \left(1+\frac{i k}{2}\right)  e^{-i k |X|}\biggr\},\!\!
	\end{align}
	and the function $v_k(X)$ satisfies
	\begin{align}\label{eq41}
		v_k(X)=\frac{2\epsilon_k-k^2}{2\epsilon_k+k^2} u_k(X).
	\end{align}
	Equations (\ref{eq21}) and (\ref{eq41}) apply for $k>0$, while for $k<0$, one should use the prescription (\ref{eq:inversion}).

	The normalization factor $N_k$ in Eq.~(\ref{eq:psi1ansatz}) is obtained from the requirement \cite{pitaevskii_bose-einstein_2003}
	\begin{align}\label{eq:normalization}
		N_k N_q\int_{-L/2\xi_\mu}^{L/2\xi_\mu} dX [u_k(X)u_q^*(X)-v_k(X)v_q^*(X)]=\delta_{k,q}.
	\end{align}
	In the case $L\gg\xi_\mu$ the normalization factor is given by \cite{petkovic_local_2022} 
	\begin{align}\label{eq:Nk}
		N_k=(\xi_\mu/2L\epsilon_k)^{1/2}.
	\end{align}
	Indeed, at $L\gg\xi_\mu$ it is sufficient to use $u_k(X)$ and $v_k(X)$ at $|X|\gg 1$ that are given by Eqs.~(\ref{eq21}) and (\ref{eq41}). Substituting them into the condition (\ref{eq:normalization}), one obtains Eq.~(\ref{eq:Nk}).

	\subsection{Second quantum correction}
	
	The mean-field solution (\ref{eq:psi0sol}) is time independent. Since it enters to the quantum contribution to the density as a term $\langle \psi_0(X) \hat\psi_2(X,T)\rangle +\langle \psi_0(X) \hat\psi_2^\dagger(X,T)\rangle$, for our purpose of evaluating the mean density (\ref{eq:density}) it is sufficient to consider Eq.~(\ref{eq:psi2}) averaged with respect to the ground state. Additional simplification arises as a consequence of the mean-field solution being a real function. For the purpose of the density that is real, we thus only need the real part of the expectation value of the averaged operator. Introducing 
	\begin{align}\label{eq:psi2Re}
		\psi_2(X)=\mathrm{Re} \langle\hat \psi_2(X,T)\rangle,
	\end{align}
	we obtain that it satisfies
	\begin{align}\label{eq:psi2deq}
		\left[\widehat{\mathcal{L}}_3(X) + \widetilde G \delta(X)\right] \psi_2(X)=h(X),
	\end{align}
	where the source term is given by
	\begin{align}\label{eq:h}
		h(X)=-\psi_0(X) \left[2\langle \hat\psi_1^\dagger(X,T) \hat\psi_1(X,T)\rangle+ \mathrm{Re}\langle \hat\psi_1^2(X,T)\rangle\right].	
	\end{align}
	In the left-hand side of Eq.~(\ref{eq:h}) we omitted the time dependence. This is a consequence of the form (\ref{eq:psi1ansatz}) for $\hat\psi_1(X,T)$, yielding the time independent expression
	\begin{align}\label{eq:h(Y)}
		h(X)={}&\frac{\psi_0(X)}{2}\sum_{k;|k|>\lambda} N_k^2\biggl[u_k(X)v_k^*(X)+u_k^*(X)v_k(X)\notag\\
		&-4|v_k(X)|^2\biggr].
	\end{align}
	Note that the  summation over real $k$ must be performed carefully, taking into account the condition $|k|>\lambda$, where $\lambda>0$ is a small cutoff. We have verified that the summand diverges as $1/k$ at $k\to 0$. Another comment concerns the time independence of the source term $h(Y)$, which enabled us to introduce the form of Eq.~(\ref{eq:psi2Re}), since $\langle\partial_T\hat\psi_2(X,T)\rangle=0$ in this case \cite{reichert_casimir-like_2019}.
	
	The expression for $\psi_2(X)$ valid at arbitrary $X$ is complicated and given in Appendix \ref{appendixA}. Instead of its analysis, let us find $\psi_2(X)$
	at $X\gg 1$ using an alternative way. To do this we notice that $\psi_2(X)$ satisfies the exact relation
	\begin{align}\label{eq:psi2X>0}
		&\frac{d}{dX}\psi_2(X) +2\tanh(X+X_0)\psi_2(X)\notag\\
		&=2\cosh^2(X+X_0) \int_X^{\infty} dY \frac{h(Y)}{\cosh^2(Y+X_0)}
	\end{align}
	for $X>0$. Indeed, acting on Eq.~(\ref{eq:psi2X>0}) with the operator $d/dX$, one directly shows that it reduces to Eq.~(\ref{eq:psi2deq}) in the case $X>0$ [i.e., to Eq.~(\ref{eq:psi2deq}) without the $\delta$-function term]. At $X\gg 1$, 
	the integrand in the right-hand side of Eq.~(\ref{eq:psi2X>0}) further simplifies as one can use
	\begin{align}
		\frac{1}{\cosh^2(Y+X_0)}=4e^{-2(Y+X_0)}+ O\left(e^{-4Y}\right).
	\end{align}	
	Equations (\ref{eq21}) and (\ref{eq41}) enable us to calculate $h(Y)$ of Eq.~(\ref{eq:h(Y)}) at $Y\gg 1$, which becomes a sum of exponential functions of $Y$. The resulting integral in Eq.~(\ref{eq:psi2X>0}) can then be easily evaluated. Finally, solving the  differential equation (\ref{eq:psi2X>0}) where in the left-hand side we approximate $\tanh(X+X_0)$ by $1$, we find 
	\begin{align}\label{eq:psi2X>>1}
		\psi_2(X)={}&\sum_{k>\lambda} N_k^2 \left(k\sqrt{4+k^2}-k^2-1\right)\times \Biggl[1 \notag\\
		&-\frac{r(k)}{2} \frac{k+2i}{k-2i} \frac{e^{2i k X}}{k^2+1}  -\frac{r^*(k)}{2} \frac{k-2i}{k+2i} \frac{e^{-2i k X}}{k^2+1}\Biggr]\notag\\
		&+O\left(e^{-2X}\right),
	\end{align}
	which is a solution at $X\gg 1$. The term containing the sum in the right-hand side of Eq.~(\ref{eq:psi2X>>1}) arises as a particular solution of the differential equation and does not have the integration constant. On the other hand, the integration constant affects the second term. However, it is exponentially suppressed and thus very small at $X\gg 1$.
	
	\section{The density profile of the Bose gas}
	
	The results of the previous section enable us to calculate the density profile of a weakly-interacting Bose gas in the presence of an impurity coupled to the system by an arbitrary coupling strength. Using Eq.~(\ref{eq:Psipsi}) and the series (\ref{eq:psiseries}), the local density of the Bose gas (\ref{eq:density}) can be expressed as
	\begin{align}\label{eq:n(x)old}
		n(x)={}&\frac{\mu}{g}\biggl\{|\psi_0(X)|^2+\alpha^2 \bigl[\langle \hat\psi_1^\dagger(X,T) \hat\psi_1(X,T)\rangle\notag\\
		&+2\psi_0(X)\psi_2(X)\bigr]+O(\alpha^3)\biggr\}_{X=x/\xi_\mu}.
	\end{align}
	The density expressed as in Eq.~(\ref{eq:n(x)old}) depends on the chemical potential of the Bose gas $\mu$ that was kept fixed in the equations of motion [see Eq.~(\ref{eq:eom})]. It is more convenient to eliminate $\mu$ and instead use the mean density $n_0$ in $n(x)$. In addition, instead of $x/\xi_\mu$ that depends on $\mu$, we will express the density in terms of $x/\xi$, where $\xi=\hbar/\sqrt{mgn_0}$ denotes the healing length. We thus need the expression for the chemical potential of a weakly-interacting Bose gas that is given by \cite{popov_theory_1977} $\mu=gn_0[1-\sqrt\gamma/\pi+O(\gamma)]$. After expressing $\mu$ in terms of $n_0$, Eq.~(\ref{eq:n(x)old}) takes the form \footnote{Below this point the variable $X$ does not have the meaning as in Eq.~(\ref{eq:X}), but rather as $x/\xi$. For simplicity we decided to keep the same notation, especially since the notion of $X$ is mathematically irrelevant in many expressions apart from its physical importance for Eq.~(\ref{eq:n(x)}).}
	\begin{align}\label{eq:n(x)}
		n(x)=n_0\left[n^{(0)}(X)+\sqrt\gamma\;\! n^{(1)}(X)+O(\gamma)\right]_{X=x/\xi}.
	\end{align}
	The first term in Eq.~(\ref{eq:n(x)}) is given by
	\begin{align}
		n^{(0)}(X)=|\psi_0(X)|^2=\tanh^2(|X|+X_0),
	\end{align}
	and represents the mean-field contribution. The shift $X_0$ is related to the impurity coupling by the relation (\ref{eq:psi0solb}). The second term in Eq.~(\ref{eq:n(x)}) has the form
	\begin{align}\label{eq:n1(x)}
		n^{(1)}(X)={}&\langle \hat\psi_1^\dagger(X,T) \hat\psi_1(X,T)\rangle+2\psi_0(X)\psi_2(X)\notag\\
		&-\frac{1}{\pi}|\psi_0(X)|^2-\frac{X}{2\pi}\frac{d}{dX}|\psi_0(X)|^2.
	\end{align}
	The latter expression with $\psi_0$ and $\hat\psi_1$ given, respectively, by Eqs.~(\ref{eq:psi0sol}), (\ref{eq:psi1ansatz}), and the expression for $\psi_2(X)$ given in Appendix \ref{appendixA} defines the quantum contribution to the density $n^{(1)}(X)$ at all distances. Taking the average value, Eq.~(\ref{eq:n1(x)}) becomes
	\begin{align}\label{eq:n(1)exp}
		n^{(1)}(X)={}&2\psi_0(X)\psi_2(X)-\frac{1}{\pi}|\psi_0(X)|^2-\frac{X}{2\pi}\frac{d}{dX}|\psi_0(X)|^2\notag\\
		&+\sum_{k>\lambda} N_k^2(|v_k(X)|^2+|v_k(-X)|^2).
	\end{align}
	Note that unlike the terms $\langle \hat\psi_1^\dagger(X,T) \hat\psi_1(X,T)\rangle$ and $2\psi_0(X)\psi_2(X)$ in Eq.~(\ref{eq:n1(x)}) that both diverge at small $k$, thus requiring the small cutoff $\lambda$, see Eq.~(\ref{eq:psi2X>>1}), their sum is finite. Therefore, in the right-hand side of Eq.~(\ref{eq:n(1)exp}), one can set $\lambda=0$ once all the terms are grouped under the same sum. This is physically expected since the density is an observable quantity that should not have any divergence in our system.

	Analytical evaluation of Eq.~(\ref{eq:n(1)exp}) for an arbitrary $X$, which reduces to the integration of elementary functions, is rather complicated. One should first perform the integration over $Y$ in the expression for $\psi_2(X)$ given in Appendix \ref{appendixA}, followed by the final summation over $k$ (that can be converted to the integral). The actual procedure was earlier performed for the special case $\widetilde{G}\to \infty$ in Ref.~\cite{petkovic_density_2020}. There was shown that $n^{(1)}(X)$ has fundamentally important contributions only at $X\gg 1$ as it is a long-ranged function. Since the mean-field contribution at $X\gg 1$ rapidly reaches the constant, the leading-order result for the density deviation $n_0-n(X)$ is controlled by $n^{(1)}(X)$. At $X\lesssim 1$, the quantum contribution $n^{(1)}(X)$, being multiplied by small $\sqrt\gamma$, only contains a small correction to the mean-field result.

	Following these remarks, let us evaluate Eq.~(\ref{eq:n(1)exp}) at $X\gg 1$ considering the case of arbitrary $\widetilde{G}$. Substitution of Eqs.~(\ref{eq21})--(\ref{eq:psi2X>>1}) yields
	\begin{align}\label{eq:n(1)X>>1}
		n^{(1)}(X)=&{}-\Biggl[\int_0^{\infty} \frac{dk}{4\pi} r(k) \frac{k+2i}{k-2i} \left(\frac{1-k^2}{1+k^2}+\frac{k}{\sqrt{4+k^2}}\right)\notag\\
		&\times e^{2ikX}+\mathrm{c.c.}\Biggr].
	\end{align}
	Here $\mathrm{c.c.}$ denotes a term that is complex conjugate to the integral. Equation (\ref{eq:n(1)X>>1}) represents the quantum contribution to the mean density at distances $X\gg 1$. Note that the constant term $-1/\pi$ that arises from the second term of the right-hand side of Eq.~(\ref{eq:n(1)exp}) is canceled by an elementary integral over $k$. The remaining terms are given by Eq.~(\ref{eq:n(1)X>>1}). 
	
	The asymptotic series expansion for the integral in Eq.~(\ref{eq:n(1)X>>1}) at $X\gg 1$ can be evaluated by the method of partial integration. For one part of it, $\int_0^{\infty} \frac{dk}{4\pi} r(k) \frac{k+2i}{k-2i} \frac{1-k^2}{1+k^2}e^{2ikX}+\mathrm{c.c.}$, we have found that it does not have the power law expansion. More precisely, we have obtained zero coefficient in front of all powers of $1/X$; we have checked numerically and obtained that the integral is proportional to $e^{-2X}$, which explains the absence of the asymptotic series. The remaining part of Eq.~(\ref{eq:n(1)X>>1})  contains the square root in the integrand and has power law contributions. We found
	\begin{align}\label{eq:integral}
		&\int_0^{\infty} \frac{dk}{4\pi} r(k) \frac{k+2i}{k-2i} \frac{k}{\sqrt{4+k^2}} e^{2ikX}=\frac{r(0)}{32\pi X^2}\notag\\
		&+\frac{r(0)+i r'(0)}{32\pi X^3}+\frac{3[5r(0)+8i r'(0)-4r''(0)]}{512\pi X^4}+O(X^{-5}).
	\end{align}
	The integral (\ref{eq:integral}) does not diverge at large values of $k$ since $r(k)$ decays sufficiently rapidly in this limit. At $X\gg 1$, it is thus controlled by the reflection amplitude (\ref{eq:r}) and its derivatives at $k\to 0$. They are given by
	\begin{gather}
		r(0)=0,\\
		r'(0)=-\frac{i}{2}\left(\widetilde{G}+\frac{2+\widetilde{G}^2}{\sqrt{4+\widetilde{G}^2}}-1\right),\\
		r''(0)={(\widetilde{G}-1)}\left(\widetilde{G}+\sqrt{4+\widetilde{G}^2} \right)- \frac{2}{4+\widetilde{G}^2}+\frac{5}{2}.
	\end{gather}
	Therefore, for the leading-order result at $X\gg 1$ we obtain
	\begin{align}\label{eq:n(1)X>>1final}
		n^{(1)}(X)=-\frac{1}{32\pi }\left(\widetilde{G}+\frac{2+\widetilde{G}^2}{\sqrt{4+\widetilde{G}^2}}-1\right)\frac{1}{|X|^3},
	\end{align}
	where at the very end we have introduced the absolute value since $n^{(1)}(X)$ is an even function that we studied so far for positive $X$. Equation (\ref{eq:n(1)X>>1final}) applies for arbitrary $\widetilde{G}$ and should be understood as the first term of the asymptotic expansion of $n^{(1)}(X)$ at $|X|\gg 1$. The mathematical divergence that occurs at $\widetilde{G}\to\infty$ leads to the infinite result, which is unphysical. One should have in mind that $X$ is assumed to be the largest parameter in Eq.~(\ref{eq:n(1)X>>1final}) and thus the operation $\widetilde{G}\to\infty$ is not allowed there. However,  $n^{(1)}(X)$ can be easily calculated at very large $\widetilde{G}$. To do this it suffices first to take the limit $\widetilde{G}\to\infty$, which imposes $r(k)=1$. Substituting this in Eq.~(\ref{eq:integral}) one obtains the leading-order result
	\begin{align}\label{eq:n(1)X>>1Ginf}
		n^{(1)}(X)=-\frac{1}{16\pi X^2}.
	\end{align} 
	A comparison of Eqs.~(\ref{eq:n(1)X>>1final}) and (\ref{eq:n(1)X>>1Ginf}) shows that at $\widetilde{G}\gg 1$, there is a crossover at distances  $X\sim \widetilde{G}$. At $1\ll X\ll \widetilde{G}$, we obtain the result (\ref{eq:n(1)X>>1Ginf}), while at $1\ll \widetilde{G}\ll X$, Eq.~(\ref{eq:n(1)X>>1final}) applies. Interested reader can easily evaluate the subleading power law corrections to Eqs.~(\ref{eq:n(1)X>>1final}) and (\ref{eq:n(1)X>>1Ginf}) using Eq.~(\ref{eq:integral}). We notice that  in $n^{(1)}(X)$ we have neglected the terms that decay exponentially, since they are always subleading. We also notice that the obtained result (\ref{eq:n(1)X>>1Ginf}) with the corresponding corrections is in agreement with the result of Ref.~\cite{petkovic_density_2020}, where $n^{(1)}(X)$ is calculated analytically for arbitrary $X$.
	
	The crossover between the results (\ref{eq:n(1)X>>1final}) and (\ref{eq:n(1)X>>1Ginf}) can be described analytically. To do so we need the expression for the reflection amplitude (\ref{eq:r}) at $k\sim 1/\widetilde{G}\ll 1$ that is given by
	\begin{align}\label{eq:rcross}
		r(k)=\frac{k}{k+i/\widetilde{G}}.
	\end{align}
	Equation (\ref{eq:rcross}) interpolates between $r(k)=-i \widetilde{G} k$ at $k\ll1/\widetilde{G}$ and $r(k)=1$ at $\widetilde{G}\gg 1/k$. Then we can find the result
	\begin{align}\label{eq:crossover}
		n^{(1)}(X)={}&-\frac{1}{16\pi}\frac{\partial^2}{\partial X^2}\int_0^{\infty} dy \frac{e^{-2X y}}{y+1/\widetilde{G}}\notag\\
		={}&-\frac{1}{16\pi X^2}+\frac{1}{8\pi \widetilde{G} X}+\frac{e^{2X/\widetilde{G}}\mathrm{Ei}(-2X/\widetilde{G})}{4\pi\widetilde{G}^2}.
	\end{align}
	Here $X>0$ and $\mathrm{Ei}(x)=-\int_{-x}^{\infty}dt e^{-t}/t$. At $X\gg \widetilde{G}$, Eq.~(\ref{eq:crossover}) becomes $n^{(1)}(X)=-\widetilde{G}/16\pi X^3$, while in the opposite case $\widetilde{G}\gg X$, it is given by $n^{(1)}(X)=-1/16\pi X^2$. This is in agreement with Eqs.~(\ref{eq:n(1)X>>1final}) and (\ref{eq:n(1)X>>1Ginf}).
	
	\section{Discussion}
	
	Being perturbed by the presence of an impurity, the local density $n(x)$ of a weakly-interacting Bose gas reaches its mean thermodynamic-limit value $n_0$ at distances $|x|\to\infty$. Within the mean-field approximation, the deviation of the density at large separations from the impurity is an exponentially small function of the distance, 
	\begin{align}
		n_0-n(x)=\frac{4\widetilde{G}n_0}{2+\sqrt{4+\widetilde{G}^2}}e^{-2|x|/\xi}.
	\end{align}
	This means that in practice, the local density $n(x)$ saturates very quickly to its mean value, which occurs at distances $x$ of several healing lengths $\xi$, see Fig.~\ref{fig}. This scenario is drastically changed once the effect of quantum fluctuations is taken into account. In this case the density deviation saturates in a much slower fashion, $n_0-n(x)=-n_0\sqrt\gamma\;\! n^{(1)}(x/\xi)$ at $|x|\gg \xi$. The functional dependence $n^{(1)}(x/\xi)$ is calculated analytically in this paper, see Eqs.~(\ref{eq:n(1)X>>1final}), (\ref{eq:n(1)X>>1Ginf}), and Eq.~(\ref{eq:n(1)X>>1}) for a more general expression. It has the power law scaling behavior with the leading-order terms
	\begin{align}\label{eq:n1summary}
		n^{(1)}(x/\xi)\propto \begin{cases}
			{\xi^2}/{x^2},\quad &\xi\ll|x|\ll \widetilde{G}\xi, \\
			{\xi^3}/{|x|^3},\quad &\widetilde{G}\xi\ll |x|,
		\end{cases}
	\end{align}
	in the regime of strongly-coupled impurity, $\widetilde{G}\gg 1$. In this case the impurity coupling imposes a new lengthscale in the system, $\widetilde G \xi$, which is much longer that the healing length $\xi$. Between the two lengths there is a wide region where the density deviation shows a decay with the universal, impurity-independent amplitude, see Eq.~(\ref{eq:n(1)X>>1Ginf}). In the limiting case of $\widetilde{G}\to\infty$, we notice the suppression of the regime with inverse cubic scaling. On the contrary, for an impurity weakly-coupled to the Bose gas, $\widetilde G\ll 1$, only the latter regime with $n^{(1)}(x/\xi)\propto \xi^3/|x|^3$ is realized.
	
	Why should one be interested in studying the deviation of the local density from its mean value beyond the mean-field result? First, this is a fundamental question and should be understood. Second, $n^{(1)}(x/\xi)$ is physically significant as it controls the induced interaction between impurities in the Bose gas, as we will  derive now. Consider our original system consisting of a Bose gas with the impurity at the origin that is locally coupled to the gas described by the Hamiltonian $\hat H$ of Eq.~(\ref{eq:H}), and let us introduce another impurity at the position $x$. The new system is described by the Hamiltonian
	\begin{align}\label{eq:HH}
		\mathcal{\hat H}=\hat H+ G_1 \hat\Psi^\dagger(x,t)\hat\Psi(x,t).
	\end{align}
	Applying the Hellmann–Feynman theorem, $\langle{\partial \mathcal{\hat H}}/{\partial G_1}\rangle_{\mathcal{\hat H}}={\partial E}/{\partial G_1}$, we obtain the relation 
	\begin{align}\label{eq:HF}
		\bigl \langle \hat\Psi^\dagger(x,t)\hat\Psi(x,t)\bigr\rangle_{\mathcal{\hat H}} =\frac{\partial E}{\partial G_1},
	\end{align}
	which connects the local density for the system described by the Hamiltonian (\ref{eq:HH}) and the derivative of its ground-state energy $E$. By $\langle \cdots\rangle_{\mathcal{\hat H}}$ we denoted the average with respect to the ground state of the Hamiltonian $\mathcal{\hat H}$. For the impurity at the position $x$ that is weakly coupled to the gas, the left-hand side of Eq.~(\ref{eq:HF}) can be understood as the local density [cf.~Eq~(\ref{eq:n(x)})] for the system described by the Hamiltonian $\hat H$. Indeed, the effect of the newly added impurity with weak coupling $G_1$ on the density depletion is negligible. The derivative on right-hand side of Eq.~(\ref{eq:HF}) nontrivialy acts only to the part of the ground-state energy that depends on $G_1$. It involves two terms. One is $x$ independent and represents the impurity binding energy and the other is Bose-medium induced inter-impurity interaction $U(x)$. The later can be extracted by integrating Eq.~(\ref{eq:HF}) and is given by $U(x)=G_1\langle n(x)\rangle -G_1 \langle n(x\to\infty)\rangle$,
	where we have chosen the integration constant to obtain physically expected absence of the inter-impurity interaction at infinite separations. Since $\langle n(x\to\infty)\rangle=n_0$, we arrive at the expression
	\begin{align}\label{eq:U}
		U(x)=U^{\C}(x)+U^{\Q}(x),
	\end{align}	
	where the classical $U^{\C}(x)$ and the quantum $U^{\Q}(x)$ contributions are given by
	\begin{align}	\label{eq:Uc}
		U^{\C}(x)={}&G_1n_0\bigl[n^{(0)}(x/\xi)-1\bigr],\\
		\label{eq:Uq}
		U^{\Q}(x)={}&G_1 n_0 \sqrt\gamma \;\! n^{(1)}(x/\xi).
	\end{align}
	At short separations, $x\lesssim\xi $, the mean-field term $n^{(0)}(x/\xi)$ in the local density (\ref{eq:n(x)}) makes dominant the classical contribution (\ref{eq:Uc}) in the induced interaction, $U(x)\approx U^{\C}(x)$. However at larger distances, $x\gg \xi$, the quantum contribution to the density $n^{(1)}(x/\xi)$ prevails, which makes the quantum contribution (\ref{eq:Uq}) dominant, $U(x)\approx U^{\Q}(x)$. 
	
	In several special cases, Eqs.~(\ref{eq:U})--(\ref{eq:Uq}) reduce to the known results. For example, at weak and equal coupling constants, $G_1=G\ll g/\sqrt\gamma$, Eqs.~(\ref{eq:Uc}) and (\ref{eq:Uq}) reduce, respectively, to the results (3) and (4) of Ref.~\cite{reichert_casimir-like_2019}. In the case of induced interaction between an infinitely strongly coupled impurity to the gas, $\widetilde{G}\to\infty$, and another one that is weakly coupled, Eqs.~(\ref{eq:U})--(\ref{eq:Uq}) become the results (22) and (23) of Ref.~\cite{reichert_fluctuation-induced_2019}. On the other hand, the present results (\ref{eq:U})--(\ref{eq:Uq}) are more general as they describe all the regimes with arbitrary coupling $G$ and weak $G_1$.

	In this paper, a weakly-interacting Bose gas is treated by a method that leads to the quasiparticle excitation spectrum described by the Bogoliubov dispersion (\ref{eq:BS}). This picture is valid at (dimensionful) momenta higher than $\hbar \gamma^{1/4}/\xi$ \cite{pustilnik_low-energy_2014}. Namely, at lowest momenta, the dispersion (\ref{eq:BS}) and the true quasiparticle dispersion agree only in the leading order; the subleading term of Eq.~(\ref{eq:BS}) starts with $|k|^3$, unlike the true dispersion that has $k^2$ term. Therefore, at distances much larger than $\xi \gamma^{-1/4}$ the power-law decay of $n^{(1)}(x/\xi)$, see Eq.~(\ref{eq:n1summary}), might be affected. At such longest distances, however, the quantum contribution of the induced interaction between the impurities is known to behaves as a power law, $U^{\Q}(x)=-m v^2 \Gamma(G)\Gamma(G_1)\xi^3/32\pi |x|^3$ \cite{schecter_phonon-mediated_2014}. Here $v=\pi \hbar n_0/mK$ is the sound velocity and the dimensionless parameter $\Gamma(G)$ denotes so-called impurity-phonon scattering amplitude. At weak coupling we have $\Gamma(G)=-G/\hbar v$ \cite{schecter_phonon-mediated_2014}. Using Eq.~(\ref{eq:Uq}), we then find the expression for $n^{(1)}(x/\xi)$ that is in agreement with our result (\ref{eq:n(1)X>>1final}) taken at the leading order in $\widetilde{G}\ll 1$. The case of strong impurity coupling we cannot discuss since we do not know $\Gamma(G)$.
	
	The conventional treatment of low-energy properties of one-dimensional quantum liquids is via the Luttinger liquid theory. The system described by the Hamiltonian (\ref{eq:H}) can be understood as a quantum liquid and thus one may be tempted to use the latter theory. It generically predicts the Friedel oscillations of the particle density around its mean with the envelop function behaving as $|x|^{-K}$ at distances from the impurity longer than $1/n_0$ \cite{cazalilla_low-energy_2002}. Here the Luttinger liquid parameter for weakly-interacting bosons is given by $K=\pi/\sqrt\gamma\gg 1$. Such rapid saturation of density oscillations is negligibly small effect in weakly-interacting Bose gases. On the contrary, we found that the local density $n(x)$ slowly approaches the mean density $n_0$, see for example Eqs.~(\ref{eq:n(x)}) and (\ref{eq:n1summary}). This should be contrasted with the Friedel oscillations that exist in the case of a strongly-interacting Bose gas. The latter system has an effective description in terms of weakly-attractive fermions. In the limit $\widetilde{G},\gamma\to\infty$, one finds the exact result \cite{sen_fermionic_2003}
	\begin{align}\label{eq:ff}
		n(x)=n_0-\frac{\sin(2\pi n_0 x)}{2\pi x}.
	\end{align}
	In the present case we have $K=1$ and thus the envelop of the density decay in Eq.~(\ref{eq:ff}) behaves according to the generic Luttinger liquid prediction. The disappearance of the oscillations in the density (\ref{eq:ff}) as the value of $\gamma$ is decreased and the formation of the hole in the density near the impurity drawn in Fig.~\ref{fig} would be interesting to study.

	This paper calls for a study of the density depletion decay in higher-dimensional Bose systems with an impurity. A phase-space argument gives the expectation for the induced inter-impurity interaction $U(x)\propto 1/|x|^{2d+1}$ at longest distances in a $d$-dimensional system \cite{schecter_phonon-mediated_2014}, which is in agreement with the detailed calculation of Ref.~\cite{fujii_universal_2022} performed for $d=3$. Our previous argument based on the Hellmann-Feynman theorem implies the same decay law for the density depletion as the one for $U(x)$. More detailed consideration of this problem is left for a future work. Another interesting direction for future research is the study of the density of an interacting system of bosons in one dimension using the techniques of integrable models. In the case $\widetilde{G}\to+\infty$, the system describes interacting bosons in the box potential, which admits an exact solution. We are however not aware of the study of the density profile of the latter system at arbitrary interaction. 
	
	\appendix
	
	\section{Solution of Eq.~(\ref{eq:psi2deq})\label{appendixA}}
	
	Equation (\ref{eq:psi2deq}) is a linear second order differential equation with a singular term at the origin. We consider it at $X\neq 0$. Denote
	\begin{align}
		\psi_2(X)=\begin{cases}
			\psi_2^{-}(X),\quad -\infty<X<0\\
			\psi_2^{+}(X),\quad 0<X<+\infty
		\end{cases}.
	\end{align}
	Equation (\ref{eq:psi2deq}) then takes the form
	\begin{align}\label{eq:inh}
		\widehat{\mathcal{L}}_3^{\pm}(X) \psi_2^{\pm}(X)=h(X),
	\end{align}
	where 
	\begin{align}
		\widehat{\mathcal{L}}_3^{\pm}(X)=-\frac{\partial_X^2}{2}+3\tanh^2(X\pm X_0)-1.
	\end{align}
	
	The homogeneous equation corresponding to Eq.~(\ref{eq:inh}) is defined by setting $h(X)=0$. This second order equation has two solutions that are given by
	\begin{subequations}\label{eq:zeromodes}
		\begin{gather}
			U^\pm(X)=\sech^2(X\pm X_0),\\
			V(X)^\pm=\sech^2(X\pm X_0) g(X\pm X_0),
		\end{gather}
	\end{subequations}
	where
	\begin{align}	
		g(X)=\frac{12X+8\sinh(2X)+\sinh(4X)}{16}.
	\end{align}
	Note that $g'(X)=2\cosh^4 X$.
	
	By the method of variation of parameters, we can construct the solution of Eq.~(\ref{eq:inh}):
	\begin{align}
		\psi_2^{-}(X)=\left[c_3-\int_X^0 dY V^-(Y)h(Y)\right] U^-(X)\notag\\
		+\left[c_4+\int_X^0 dY U^-(Y)h(Y)\right] V^-(X),\\
		\label{eq:psi2+}
		\psi_2^{+}(X)=\left[c_1+\int_0^X dY V^+(Y)h(Y)\right] U^+(X)\notag\\
		+\left[c_2-\int_0^X dY U^+(Y)h(Y)\right] V^+(X).
	\end{align}
	Here $c_1$, $c_2$, $c_3$, and $c_4$ are the constants that are to be determined from boundary conditions. Noticing that $V^{\pm}(\pm\infty)$ is divergent and requiring finite value of $\psi_{2}^\pm(\pm\infty)$ we are able to set
	\begin{align}\label{eq:c2}
		c_2=\int_0^{+\infty} dY U^+(Y)h(Y),\\
		c_4=-\int_{-\infty}^0 dY U^-(Y)h(Y).
	\end{align}
	The remaining two constants are determined by the boundary conditions
	\begin{gather}
		\psi_2^-(0)=\psi_2^+(0),\\
		\left(\frac{d\psi_2^+(X)}{dX}-	\frac{d\psi_2^-(X)}{dX}\right)\bigg{|}_{X=0}=2\widetilde G \psi_{2}^+(0).
	\end{gather}
	This leads to
	\begin{align}
		c_1={}&-g(X_0)\int_0^{+\infty} dY U^{+}(Y)h(Y)\notag\\
		&+\frac{g'(X_0)}{2(\widetilde{G}+2\tanh X_0)} \int_{-\infty}^{+\infty} U(Y)h(Y)\\
		c_3={}&-g(X_0)\int_{-\infty}^0 dY U^{-}(Y)h(Y)\notag\\
		&+\frac{g'(X_0)}{2(\widetilde{G}+2\tanh X_0)} \int_{-\infty}^{+\infty} U(Y)h(Y).
	\end{align}
	Here $U(Y)=U^{-}(Y)$ for $Y<0$ and $U(Y)=U^{+}(Y)$ for $Y>0$. 
	
	The latter solution simplifies in our case, since $h(Y)$ is an even function, $h(Y)=h(-Y)$, which follows from Eqs.~(\ref{eq:h(Y)}) and (\ref{eq:inversion}).
	Then we have two independent constants, e.g., $c_1$ and $c_2$, since $c_3=c_1$ and $c_4=-c_2$. Moreover, $c_1$ simplifies and becomes
	\begin{align}\label{eq:c1}
		c_1=\frac{1}{4}\left(\frac{4+3\widetilde G^2}{\widetilde G^2\sqrt{4+\widetilde G^2}}-3X_0\right) \int_0^{+\infty} dY U^{+}(Y)h(Y).
	\end{align}
	Substitution of Eqs.~(\ref{eq:c1}) and (\ref{eq:c2}) into Eq.~(\ref{eq:psi2+}) gives $\psi_2(X)$, which is the solution of Eq.~(\ref{eq:psi2deq}) at $X>0$. The solution at $X<0$ follows from the parity of $\psi_2(X)$.
	

\begin{thebibliography}{17}%
		\makeatletter
		\providecommand \@ifxundefined [1]{%
			\@ifx{#1\undefined}
		}%
		\providecommand \@ifnum [1]{%
			\ifnum #1\expandafter \@firstoftwo
			\else \expandafter \@secondoftwo
			\fi
		}%
		\providecommand \@ifx [1]{%
			\ifx #1\expandafter \@firstoftwo
			\else \expandafter \@secondoftwo
			\fi
		}%
		\providecommand \natexlab [1]{#1}%
		\providecommand \enquote  [1]{``#1''}%
		\providecommand \bibnamefont  [1]{#1}%
		\providecommand \bibfnamefont [1]{#1}%
		\providecommand \citenamefont [1]{#1}%
		\providecommand \href@noop [0]{\@secondoftwo}%
		\providecommand \href [0]{\begingroup \@sanitize@url \@href}%
		\providecommand \@href[1]{\@@startlink{#1}\@@href}%
		\providecommand \@@href[1]{\endgroup#1\@@endlink}%
		\providecommand \@sanitize@url [0]{\catcode `\\12\catcode `\$12\catcode
			`\&12\catcode `\#12\catcode `\^12\catcode `\_12\catcode `\%12\relax}%
		\providecommand \@@startlink[1]{}%
		\providecommand \@@endlink[0]{}%
		\providecommand \url  [0]{\begingroup\@sanitize@url \@url }%
		\providecommand \@url [1]{\endgroup\@href {#1}{\urlprefix }}%
		\providecommand \urlprefix  [0]{URL }%
		\providecommand \Eprint [0]{\href }%
		\providecommand \doibase [0]{http://dx.doi.org/}%
		\providecommand \selectlanguage [0]{\@gobble}%
		\providecommand \bibinfo  [0]{\@secondoftwo}%
		\providecommand \bibfield  [0]{\@secondoftwo}%
		\providecommand \translation [1]{[#1]}%
		\providecommand \BibitemOpen [0]{}%
		\providecommand \bibitemStop [0]{}%
		\providecommand \bibitemNoStop [0]{.\EOS\space}%
		\providecommand \EOS [0]{\spacefactor3000\relax}%
		\providecommand \BibitemShut  [1]{\csname bibitem#1\endcsname}%
		\let\auto@bib@innerbib\@empty
		\bibitem [{\citenamefont {Kane}\ and\ \citenamefont
			{Fisher}(1992)}]{kane_transmission_1992}%
		\BibitemOpen
		\bibfield  {author} {\bibinfo {author} {\bibfnamefont {C.~L.}\ \bibnamefont
				{Kane}}\ and\ \bibinfo {author} {\bibfnamefont {M.~P.~A.}\ \bibnamefont
				{Fisher}},\ }\bibfield  {title} {\enquote {\bibinfo {title} {Transmission
					through barriers and resonant tunneling in an interacting one-dimensional
					electron gas},}\ }\href {\doibase 10.1103/PhysRevB.46.15233} {\bibfield
			{journal} {\bibinfo  {journal} {Phys. Rev. B}\ }\textbf {\bibinfo {volume}
				{46}},\ \bibinfo {pages} {15233} (\bibinfo {year} {1992})}\BibitemShut
		{NoStop}%
		\bibitem [{\citenamefont {Fabrizio}\ and\ \citenamefont
			{Gogolin}(1995)}]{fabrizio_interacting_1995}%
		\BibitemOpen
		\bibfield  {author} {\bibinfo {author} {\bibfnamefont {M.}~\bibnamefont
				{Fabrizio}}\ and\ \bibinfo {author} {\bibfnamefont {A.~O.}\ \bibnamefont
				{Gogolin}},\ }\bibfield  {title} {\enquote {\bibinfo {title} {Interacting
					one-dimensional electron gas with open boundaries},}\ }\href {\doibase
			10.1103/PhysRevB.51.17827} {\bibfield  {journal} {\bibinfo  {journal} {Phys.
					Rev. B}\ }\textbf {\bibinfo {volume} {51}},\ \bibinfo {pages} {17827}
			(\bibinfo {year} {1995})}\BibitemShut {NoStop}%
		\bibitem [{\citenamefont {Egger}\ and\ \citenamefont
			{Grabert}(1995)}]{egger_friedel_1995}%
		\BibitemOpen
		\bibfield  {author} {\bibinfo {author} {\bibfnamefont {R.}~\bibnamefont
				{Egger}}\ and\ \bibinfo {author} {\bibfnamefont {H.}~\bibnamefont
				{Grabert}},\ }\bibfield  {title} {\enquote {\bibinfo {title} {Friedel
					{Oscillations} for {Interacting} {Fermions} in {One} {Dimension}},}\ }\href
		{\doibase 10.1103/PhysRevLett.75.3505} {\bibfield  {journal} {\bibinfo
				{journal} {Phys. Rev. Lett.}\ }\textbf {\bibinfo {volume} {75}},\ \bibinfo
			{pages} {3505} (\bibinfo {year} {1995})}\BibitemShut {NoStop}%
		\bibitem [{Note1()}]{Note1}%
		\BibitemOpen
		\bibinfo {note} {Note that Eq.~(\ref {eq:psi0sol}) is also the nodeless
			solution of Eq.~(\ref {eq:GP}) in the case $\protect \mbox {$\setbox \z@
				\hbox {\mathsurround \z@ $\textstyle G$}\mathaccent "0365{G}$}<0$. In this
			case the solution is nonsingular (normalizable) at any attraction apart from
			the extreme case $\protect \mbox {$\setbox \z@ \hbox {\mathsurround \z@
					$\textstyle G$}\mathaccent "0365{G}$}\to -\infty $.}\BibitemShut {Stop}%
		\bibitem [{\citenamefont {Kovrizhin}(2001)}]{kovrizhin_exact_2001}%
		\BibitemOpen
		\bibfield  {author} {\bibinfo {author} {\bibfnamefont {D.~L.}\ \bibnamefont
				{Kovrizhin}},\ }\bibfield  {title} {\enquote {\bibinfo {title} {Exact form of
					the {Bogoliubov} excitations in one-dimensional nonlinear {Schrödinger}
					equation},}\ }\href {\doibase 10.1016/S0375-9601(01)00503-5} {\bibfield
			{journal} {\bibinfo  {journal} {Phys. Lett. A}\ }\textbf {\bibinfo {volume}
				{287}},\ \bibinfo {pages} {392} (\bibinfo {year} {2001})}\BibitemShut
		{NoStop}%
		\bibitem [{\citenamefont {Petković}(2022)}]{petkovic_local_2022}%
		\BibitemOpen
		\bibfield  {author} {\bibinfo {author} {\bibfnamefont {A.}~\bibnamefont
				{Petković}},\ }\bibfield  {title} {\enquote {\bibinfo {title} {Local
					spectral density of an interacting one-dimensional {Bose} gas with an
					impurity},}\ }\href {\doibase 10.1103/PhysRevA.105.043305} {\bibfield
			{journal} {\bibinfo  {journal} {Phys. Rev. A}\ }\textbf {\bibinfo {volume}
				{105}},\ \bibinfo {pages} {043305} (\bibinfo {year} {2022})}\BibitemShut
		{NoStop}%
		\bibitem [{\citenamefont {Pitaevskii}\ and\ \citenamefont
			{Stringari}(2003)}]{pitaevskii_bose-einstein_2003}%
		\BibitemOpen
		\bibfield  {author} {\bibinfo {author} {\bibfnamefont {L.~P.}\ \bibnamefont
				{Pitaevskii}}\ and\ \bibinfo {author} {\bibfnamefont {S.}~\bibnamefont
				{Stringari}},\ }\href@noop {} {\emph {\bibinfo {title} {Bose-{Einstein}
					{Condensation}}}}\ (\bibinfo  {publisher} {Oxford University Press},\
		\bibinfo {year} {2003})\BibitemShut {NoStop}%
		\bibitem [{\citenamefont {Reichert}\ \emph
			{et~al.}(2019{\natexlab{a}})\citenamefont {Reichert}, \citenamefont
			{Ristivojevic},\ and\ \citenamefont
			{Petković}}]{reichert_casimir-like_2019}%
		\BibitemOpen
		\bibfield  {author} {\bibinfo {author} {\bibfnamefont {B.}~\bibnamefont
				{Reichert}}, \bibinfo {author} {\bibfnamefont {Z.}~\bibnamefont
				{Ristivojevic}}, \ and\ \bibinfo {author} {\bibfnamefont {A.}~\bibnamefont
				{Petković}},\ }\bibfield  {title} {\enquote {\bibinfo {title} {The
					{Casimir}-like effect in a one-dimensional {Bose} gas},}\ }\href {\doibase
			10.1088/1367-2630/ab1b8e} {\bibfield  {journal} {\bibinfo  {journal} {New J.
					Phys.}\ }\textbf {\bibinfo {volume} {21}},\ \bibinfo {pages} {053024}
			(\bibinfo {year} {2019}{\natexlab{a}})}\BibitemShut {NoStop}%
		\bibitem [{\citenamefont {Popov}(1977)}]{popov_theory_1977}%
		\BibitemOpen
		\bibfield  {author} {\bibinfo {author} {\bibfnamefont {V.~N.}\ \bibnamefont
				{Popov}},\ }\bibfield  {title} {\enquote {\bibinfo {title} {Theory of
					one-dimensional {Bose} gas with point interaction},}\ }\href {\doibase
			10.1007/BF01036714} {\bibfield  {journal} {\bibinfo  {journal} {Theor. Math.
					Phys.}\ }\textbf {\bibinfo {volume} {30}},\ \bibinfo {pages} {222} (\bibinfo
			{year} {1977})}\BibitemShut {NoStop}%
		\bibitem [{Note2()}]{Note2}%
		\BibitemOpen
		\bibinfo {note} {Below this point the variable $X$ does not have the meaning
			as in Eq.~(\ref {eq:X}), but rather as $x/\xi $. For simplicity we decided to
			keep the same notation, especially since the notion of $X$ is mathematically
			irrelevant in many expressions apart from its physical importance for
			Eq.~(\ref {eq:n(x)}).}\BibitemShut {Stop}%
		\bibitem [{\citenamefont {Petković}\ \emph {et~al.}(2020)\citenamefont
			{Petković}, \citenamefont {Reichert},\ and\ \citenamefont
			{Ristivojevic}}]{petkovic_density_2020}%
		\BibitemOpen
		\bibfield  {author} {\bibinfo {author} {\bibfnamefont {A.}~\bibnamefont
				{Petković}}, \bibinfo {author} {\bibfnamefont {B.}~\bibnamefont {Reichert}},
			\ and\ \bibinfo {author} {\bibfnamefont {Z.}~\bibnamefont {Ristivojevic}},\
		}\bibfield  {title} {\enquote {\bibinfo {title} {Density profile of a
					semi-infinite one-dimensional {Bose} gas and bound states of the impurity},}\
		}\href {\doibase 10.1103/PhysRevResearch.2.043104} {\bibfield  {journal}
			{\bibinfo  {journal} {Phys. Rev. Research}\ }\textbf {\bibinfo {volume}
				{2}},\ \bibinfo {pages} {043104} (\bibinfo {year} {2020})}\BibitemShut
		{NoStop}%
		\bibitem [{\citenamefont {Reichert}\ \emph
			{et~al.}(2019{\natexlab{b}})\citenamefont {Reichert}, \citenamefont
			{Petković},\ and\ \citenamefont
			{Ristivojevic}}]{reichert_fluctuation-induced_2019}%
		\BibitemOpen
		\bibfield  {author} {\bibinfo {author} {\bibfnamefont {B.}~\bibnamefont
				{Reichert}}, \bibinfo {author} {\bibfnamefont {A.}~\bibnamefont {Petković}},
			\ and\ \bibinfo {author} {\bibfnamefont {Z.}~\bibnamefont {Ristivojevic}},\
		}\bibfield  {title} {\enquote {\bibinfo {title} {Fluctuation-induced
					potential for an impurity in a semi-infinite one-dimensional {Bose} gas},}\
		}\href {\doibase 10.1103/PhysRevB.100.235431} {\bibfield  {journal} {\bibinfo
				{journal} {Phys. Rev. B}\ }\textbf {\bibinfo {volume} {100}},\ \bibinfo
			{pages} {235431} (\bibinfo {year} {2019}{\natexlab{b}})}\BibitemShut
		{NoStop}%
		\bibitem [{\citenamefont {Pustilnik}\ and\ \citenamefont
			{Matveev}(2014)}]{pustilnik_low-energy_2014}%
		\BibitemOpen
		\bibfield  {author} {\bibinfo {author} {\bibfnamefont {M.}~\bibnamefont
				{Pustilnik}}\ and\ \bibinfo {author} {\bibfnamefont {K.~A.}\ \bibnamefont
				{Matveev}},\ }\bibfield  {title} {\enquote {\bibinfo {title} {Low-energy
					excitations of a one-dimensional {Bose} gas with weak contact repulsion},}\
		}\href {\doibase 10.1103/PhysRevB.89.100504} {\bibfield  {journal} {\bibinfo
				{journal} {Phys. Rev. B}\ }\textbf {\bibinfo {volume} {89}},\ \bibinfo
			{pages} {100504} (\bibinfo {year} {2014})}\BibitemShut {NoStop}%
		\bibitem [{\citenamefont {Schecter}\ and\ \citenamefont
			{Kamenev}(2014)}]{schecter_phonon-mediated_2014}%
		\BibitemOpen
		\bibfield  {author} {\bibinfo {author} {\bibfnamefont {M.}~\bibnamefont
				{Schecter}}\ and\ \bibinfo {author} {\bibfnamefont {A.}~\bibnamefont
				{Kamenev}},\ }\bibfield  {title} {\enquote {\bibinfo {title}
				{Phonon-{Mediated} {Casimir} {Interaction} between {Mobile} {Impurities} in
					{One}-{Dimensional} {Quantum} {Liquids}},}\ }\href {\doibase
			10.1103/PhysRevLett.112.155301} {\bibfield  {journal} {\bibinfo  {journal}
				{Phys. Rev. Lett.}\ }\textbf {\bibinfo {volume} {112}},\ \bibinfo {pages}
			{155301} (\bibinfo {year} {2014})}\BibitemShut {NoStop}%
		\bibitem [{\citenamefont {Cazalilla}(2002)}]{cazalilla_low-energy_2002}%
		\BibitemOpen
		\bibfield  {author} {\bibinfo {author} {\bibfnamefont {M.~A.}\ \bibnamefont
				{Cazalilla}},\ }\bibfield  {title} {\enquote {\bibinfo {title} {Low-energy
					properties of a one-dimensional system of interacting bosons with
					boundaries},}\ }\href {\doibase 10.1209/epl/i2002-00112-5} {\bibfield
			{journal} {\bibinfo  {journal} {Europhys. Lett.}\ }\textbf {\bibinfo {volume}
				{59}},\ \bibinfo {pages} {793} (\bibinfo {year} {2002})}\BibitemShut
		{NoStop}%
		\bibitem [{\citenamefont {Sen}(2003)}]{sen_fermionic_2003}%
		\BibitemOpen
		\bibfield  {author} {\bibinfo {author} {\bibfnamefont {D.}~\bibnamefont
				{Sen}},\ }\bibfield  {title} {\enquote {\bibinfo {title} {The fermionic limit
					of the -function {Bose} gas: a pseudopotential approach},}\ }\href {\doibase
			10.1088/0305-4470/36/27/305} {\bibfield  {journal} {\bibinfo  {journal} {J.
					Phys. A Math. Theor.}\ }\textbf {\bibinfo {volume} {36}},\ \bibinfo {pages}
			{7517} (\bibinfo {year} {2003})}\BibitemShut {NoStop}%
		\bibitem [{\citenamefont {Fujii}\ \emph {et~al.}(2022)\citenamefont {Fujii},
			\citenamefont {Hongo},\ and\ \citenamefont {Enss}}]{fujii_universal_2022}%
		\BibitemOpen
		\bibfield  {author} {\bibinfo {author} {\bibfnamefont {K.}~\bibnamefont
				{Fujii}}, \bibinfo {author} {\bibfnamefont {M.}~\bibnamefont {Hongo}}, \ and\
			\bibinfo {author} {\bibfnamefont {T.}~\bibnamefont {Enss}},\ }\bibfield
		{title} {\enquote {\bibinfo {title} {Universal van der {Waals} {Force}
					between {Heavy} {Polarons} in {Superfluids}},}\ }\href {\doibase
			10.1103/PhysRevLett.129.233401} {\bibfield  {journal} {\bibinfo  {journal}
				{Phys. Rev. Lett.}\ }\textbf {\bibinfo {volume} {129}},\ \bibinfo {pages}
			{233401} (\bibinfo {year} {2022})}\BibitemShut {NoStop}%
	\end{thebibliography}
	
	%

\end{document}